\documentclass{nature}
\usepackage{graphicx}
\usepackage{amsmath}
\usepackage{amssymb}
\usepackage{color}
\usepackage{url}
\usepackage{verbatim}
\usepackage{tabu}
\usepackage{sansmath}
\usepackage{epsfig}

\graphicspath{{images/}{./}}

\defbibheading{subbibliography}{%
  \section*{}}

\let\cite\autocite

\title{Dense magnetized plasma associated with a fast radio burst}

\author{
    {Kiyoshi~Masui}$^{1,2}$, \allowbreak
    {Hsiu-Hsien~Lin}$^{3}$, \allowbreak
    {Jonathan~Sievers}$^{4,5}$, \allowbreak
    {Christopher~J.~Anderson}$^{6}$, \allowbreak
    {Tzu-Ching~Chang}$^{7}$, \allowbreak
    {Xuelei~Chen}$^{8,9}$, \allowbreak
    {Apratim~Ganguly}$^{10}$, \allowbreak
    {Miranda~Jarvis}$^{11}$, \allowbreak
    {Cheng-Yu~Kuo}$^{12,7}$, \allowbreak
    {Yi-Chao~Li}$^{8}$, \allowbreak
    {Yu-Wei~Liao}$^{7}$, \allowbreak
    {Maura~McLaughlin}$^{13}$, \allowbreak
    {Ue-Li~Pen}$^{14,2,15}$, \allowbreak
    {Jeffrey~B.~Peterson}$^{3}$, \allowbreak
    {Alexander~Roman}$^{3}$, \allowbreak
    {Peter~T.~Timbie}$^{6}$, \allowbreak
    {Tabitha~Voytek}$^{4,3}$ \allowbreak
    \&
    {Jaswant~K.~Yadav}$^{16}$
    }

\begin{document}

\maketitle

\blfootnote{
\begin{affiliations}
    \item{Department of Physics and Astronomy, University of British Columbia, 6224 Agricultural Rd, Vancouver, BC, V6T 1Z1, Canada}
    \item{Canadian Institute for Advanced Research, CIFAR Program in Cosmology and Gravity, Toronto, ON, M5G 1Z8}
    \item{McWilliams Center for Cosmology, Carnegie Mellon University, Department of Physics, 5000 Forbes Ave, Pittsburgh, PA, 15213, USA}
    \item{Astrophysics and Cosmology Research Unit, School of Chemistry and Physics, University of KwaZulu-Natal, Durban, 4001, South Africa}
    \item{National Institute for Theoretical Physics (NITheP), KZN node, Durban, 4001, South Africa}
    \item{Department of Physics, University of Wisconsin, Madison, WI 53706-1390, USA}
    \item{Academia Sinica Institute of Astronomy and Astrophysics, 11F of Astro-Math Building, AS/NTU, 1, Sec. 4, Roosevelt Rd, Taipei 10617, Taiwan}
    \item{National Astronomical Observatories, Chinese Academy of Science, 20A Datun Road, Beijing 100012, China}
    \item{Center of High Energy Physics, Peking University, Beijing 100871, China}
    \item{Astrophysics and Cosmology Research Unit, School of Mathematics, Statistics, and Computer Science, University of KwaZulu-Natal, Durban, 4001, South Africa}
    \item{Department of Astronomy \& Astrophysics, University of Toronto, 50 St George St, Toronto, ON, M5S 3H4, Canada}
    \item{Department of Physics, National Sun Yat-Sen University No.70, Lianhai Rd., Gushan Dist., Kaohsiung City 804, Taiwan}
    \item{Department of Physics and Astronomy, West Virginia University, Morgantown WV 26506, USA}
    \item{Canadian Institute for Theoretical Astrophysics, 60 St George St, Toronto, ON, M5S 3H8, Canada}
    \item{Perimeter Institute, 31 Caroline St, Waterloo, Canada}
    \item{Indian Institute of Science Education and Research Mohali, Knowledge City, Sector 81, SAS Nagar, Manauli PO 140306, India}
\end{affiliations}
}

% Throughout we need to figure out tense: past tense (burst happened in 2011
% and is over) or present ("the burst has RM=..."). -KM

% Magnetars: starquakes vs flares vs crust disruptions? -KM

\begin{abstract}

Fast Radio Bursts are bright, unresolved, non-repeating, broadband,
millisecond flashes, found primarily at high Galactic latitudes, with dispersion measures much larger than
expected for a Galactic source\cite{Lorimer02112007,2012MNRAS.425L..71K,
2013Sci...341...53T, 2014ApJ...790..101S, 2014ApJ...792...19B,
2015MNRAS.447..246P,2015ApJ...799L...5R,2015arXiv151107746C}.
The inferred all-sky burst rate  \cite{2015arXiv150500834R} is comparable to
the core-collapse supernova rate\cite{2014ApJ...792..135T} out to redshift 0.5.
If the observed dispersion measures are assumed
to be dominated by the intergalactic medium, the sources are at cosmological
distances with redshifts\cite{2004MNRAS.348..999I,
2003ApJ...598L..79I} of 0.2 to 1.
These parameters are consistent with a wide
range of source models\cite{2014MNRAS.439L..46L,2014ApJ...797...70K,
2015ApJ...809...24G,2014MNRAS.442L...9L,2014A&A...562A.137F,2015arXiv150505535C}.
One fast radio burst \cite{2015MNRAS.447..246P} showed
circular polarization [21(7)\%] of the radio emission, but no linear polarization
was detected, and
hence no Faraday rotation measure could be determined.
Here we report the examination of archival data revealing Faraday
rotation in a newly detected burst---FRB~110523.
It has radio flux at least 0.6\,Jy and
dispersion measure 623.30(5)\,pc\,cm$^{-3}$. 
Using Galactic contribution 45\,pc\,cm$^{-3}$
and a model of intergalactic electron density\cite{2004MNRAS.348..999I},
we place the source at a maximum redshift of 0.5. 
% added by LS and not accurate...I suggest we just delete this line(JP) 
%\red{This rules out any Galactic origin for this particular source.} 
The burst has rotation measure
--186.1(1.4)\,rad\,m$^{-2}$, much higher than expected for this line of sight
through the Milky Way and the intergalactic medium, indicating magnetization in
the vicinity of the source itself or within a host galaxy. The pulse was
scattered by two distinct plasma screens during propagation, which requires
either a dense nebula associated with the source or a location within
the central region of its host galaxy. Keeping in mind that there may be more
than one type of fast radio burst source, the detection in this instance of
source-local
magnetization and scattering 
%opens the possibility that the majority of the
%dispersion is also caused by the source-local material, making the sources more
%nearby than previously assumed.
favours models
involving young stellar populations such as magnetars over models involving
the mergers of older neutron stars, which are more likely to be located in
low density regions of the host galaxy.

\end{abstract}

\begin{figure}[htbp]
\centering
\includegraphics[width=9cm]{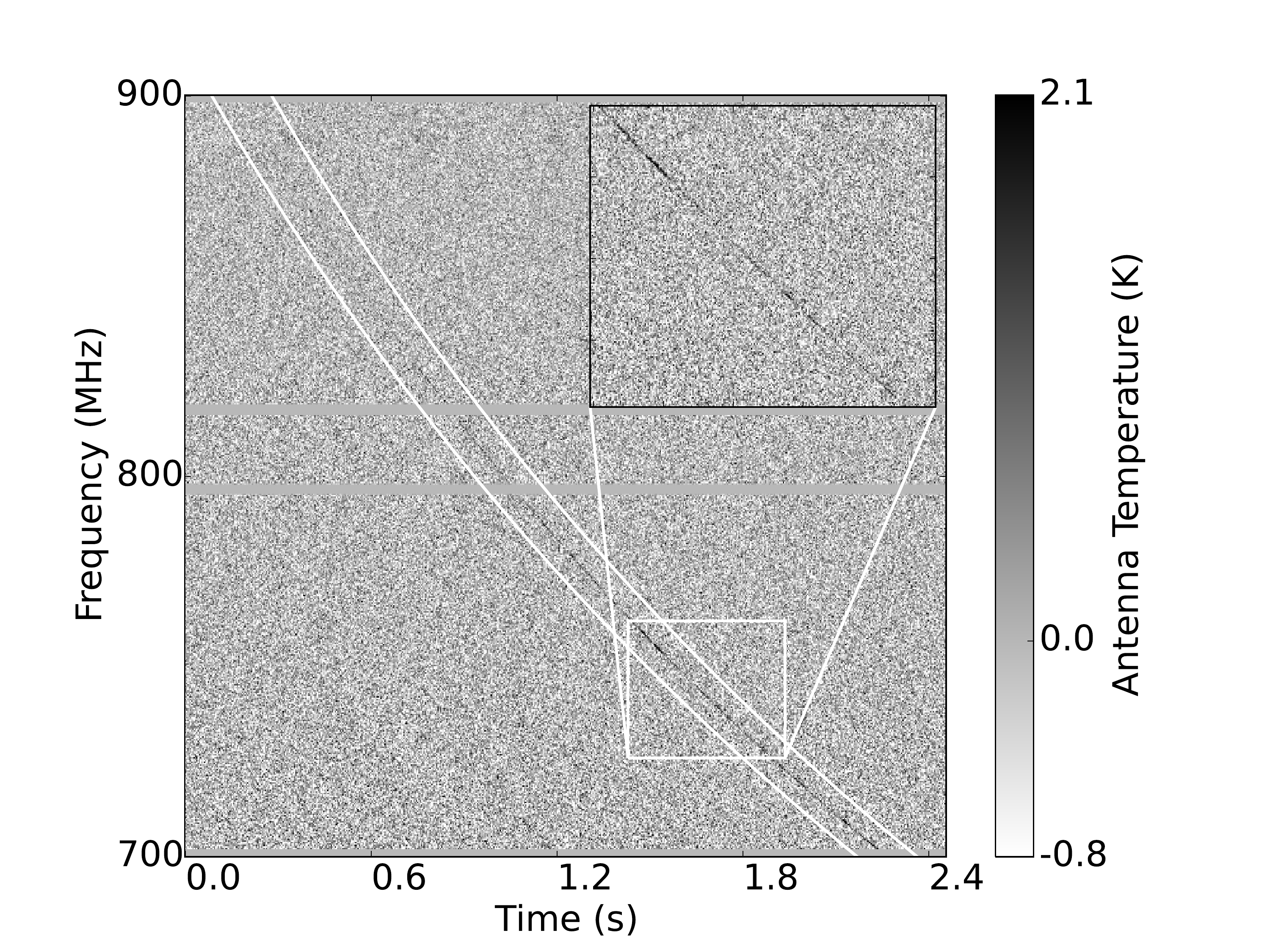} 
    \caption{
\label{fig:waterfall}
{\bf Brightness temperature spectra \textit{vs} time for FRB 110523.} The diagonal
curve shows the pulse of radio brightness sweeping in time. The arrival time is
differentially delayed (dispersed) by plasma along the line of sight. A pair
of curves in white have been added, bracketing the FRB pulse, to show that the
delay function matches that expected from cold plasma. The gray horizontal
bars show where data has been omitted due to resonances within the GBT receiver.
The inset shows fluctuations in brightness caused by scintillation. }
\end{figure}

We searched for FRBs in a data archive we collected for the Green Bank Hydrogen
Intensity Mapping
survey\cite{2010Natur.466..463C,2013ApJ...763L..20M,2013MNRAS.434L..46S}.  The
data span the frequency range 700 to 900~MHz in 4096 spectral channels.
Average spectra are recorded at 1.024~ms
intervals. We developed a new tree dedispersion algorithm and associated
computer program to search for FRBs.  First we removed cold plasma dispersion,
a frequency-dependent time delay 
$$
t_{\rm delay} = 4148.808\,{\rm s} \left(\frac{\rm DM }{\rm
pc\,cm^{-3}}\right)\left(\frac{\rm MHz^2}{\nu^{2}}\right),
$$
where $\nu$ is the radio frequency and the 
dispersion measure, ${\rm DM} = \int n_e dl$, is the line of sight integral of
the free electron number density. We then summed all frequency channels
for DM values ranging from 0 to 2000\,pc\,cm$^{-3}$ and flagged as candidates all 
data sets with eight-$\sigma$
positive excursions of flux.  These 6496 candidates were examined
by eye and compared to synthetic DM-time images of simulated FRB events.  Most
of these candidates have the characteristics of radio frequency interference
(RFI) but one matched the expected pattern of an FRB (see
Figure~\ref{fig:waterfall} and Extended Data Figure~\ref{fig:dmt}). This burst,
hereafter FRB~110523, has a DM of 623.30(6)\,pc\,cm$^{-3}$; the maximum DM
expected in this direction due to Galactic
contributions\cite{2002astro.ph..7156C} is 45\,pc\,cm$^{-3}$.
{Detailed parameters for the burst are given in Extended Data
Table~\ref{table:FRB_param_main}.}

Our detection in a
distinct band and with independent instrumentation compared to the
1.4\,GHz detections at the Parkes and Arecibo observatories greatly
strengthens the case that FRBs are astrophysical phenomena. 
In addition, as described in the Methods, the close fits to astronomical
expectations for 
dispersion spectral index, Faraday rotation, and scattering spectral index
all further support an astronomical origin.

Fitting a model to the burst data
we find the detection significance is over 40$\sigma$,
with fluence 3.79(15)\,K\,ms at our centre spectral
frequency of 800\,MHz.
The burst has a steep spectral index −-7.8(4) which we
attribute to telescope motion. 
The peak antenna temperature at 800\,MHz is 1.16(5)\,K.
We do not know the location of the source
within the GBT beam profile, but
if the source location were at beam centre where the antenna gain is 
2\,K\,Jy$^{-1}$
the measured antenna temperature would translate to 0.6\,Jy.
Off centre the antenna has lower gain so this is a
lower limit to the FRB flux, similar to that of previously
reported FRBs.
The intrinsic FWHM duration 
of the burst (with scattering removed) is $1.74(17)$\,ms, also
similar to the widths of previously reported FRBs.

Allowing the dispersion relation to vary in the model, we find that the
dispersive delay is proportional to $t_{\rm delay} \sim \nu^{-1.998(3)}$, 
in close agreement to
the expected  $\nu^{-2}$ dependence for a cold, diffuse plasma.  Following
Katz\cite{2015arXiv150506220K}, this provides
an upper limit on the density of electrons in the dispersing plasma of
$n_e < 1.3\times10^7$\,cm$^{-3}$ at 95\% confidence and a lower limit on the
size of the dispersive region of
$d > 10$\,AU. 
This limit improves upon limits from previous
bursts\cite{2014MNRAS.441L..26T,2014MNRAS.443L..11D,2015arXiv150506220K}
and rules out a flare star as the source of FRB~110523,
as stellar corona are denser and less extended by at least an order of
magnitude\cite{2015MNRAS.454.2183M}.
Flare stars being the last viable Galactic-origin model for FRB sources,
we take the source to be extragalactic.

\begin{figure}[tbp]
\centering
\includegraphics[width=9cm]{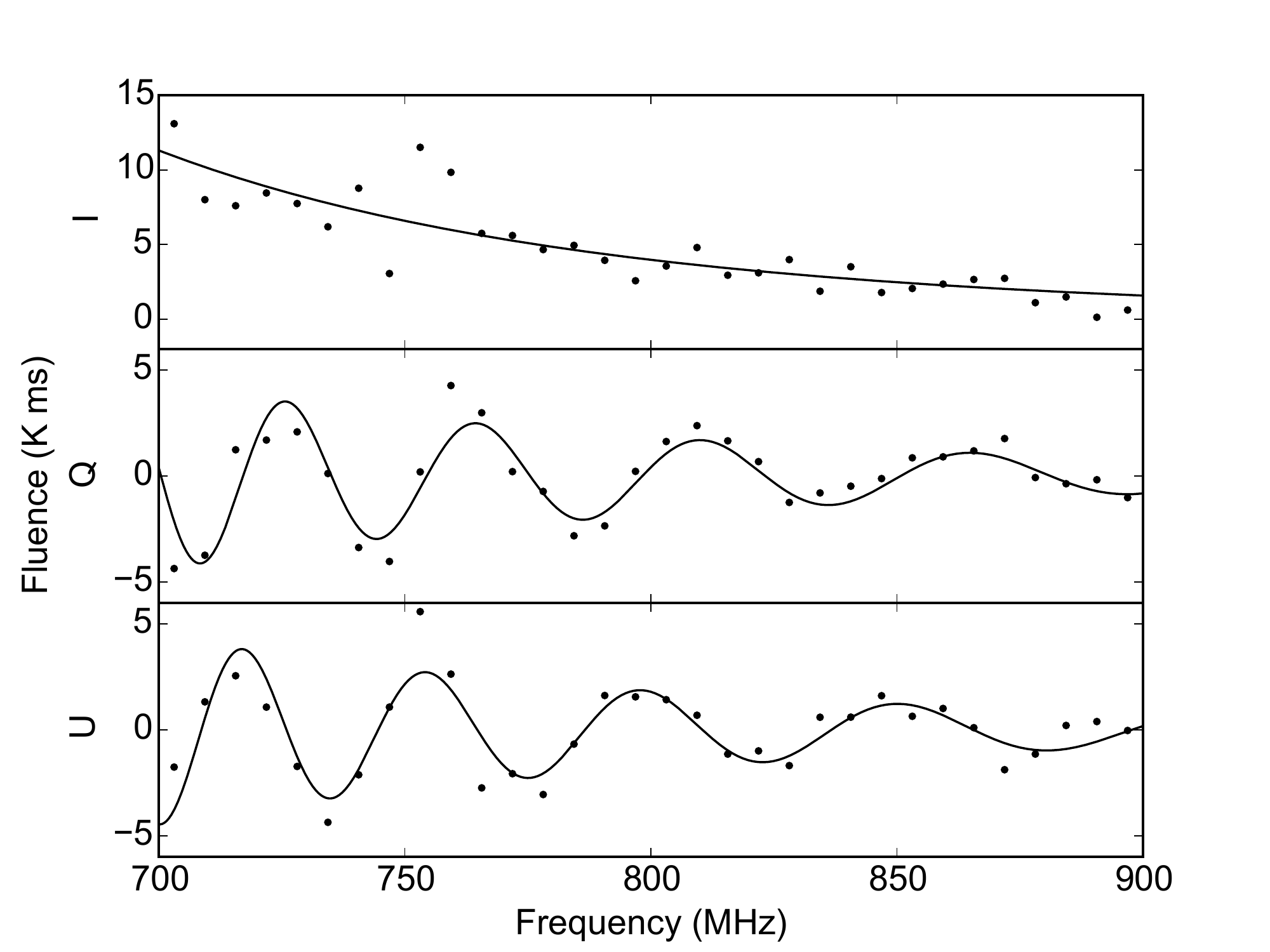}
\caption{
\label{fig:spectra}
{\bf FRB~110523 spectra in total intensity and polarization.} Plotted is the
pulse energy for total intensity (Stokes I), and linear polarization
(Stokes Q, and U). Solid curves are model fits. In addition to noise, scatter in the
measurement around the models is due to the scintillation  
visible in Figure~\ref{fig:waterfall}. 
The decline of intensity with 
frequency is primarily due to motion of the telescope beam across the sky and is
not intrinsic to the source.}
\end{figure}

We find strong
linear polarization, with linearly polarized fraction $44(3)$\%.
Linearly polarized
radio sources exhibit Faraday rotation of the polarization angle on the sky by
angle $\phi_{far}= {\rm RM} \lambda^2$, 
where $\lambda$ is the wavelength and the rotation measure, a measure of
magnetization, is the line of sight
component of the magnetic field weighted by the electron density:
$$
{\rm RM} = 0.812\,{\rm rad\,m^{-2}}
\int \frac{n_e}{\rm cm^{-3}} \frac{B_{\parallel}}{\mu {\rm G}} \frac{dl}{\rm
pc}.
$$
We detect the expected $\lambda^2$
modulation pattern in the polarization as shown in Figure~\ref{fig:spectra}.  
The best-fit RM is $-186.1(1.4)\,{\rm rad ~m}^{-2}$.
All radio telescopes have polarization leakage, an instrument-induced false
polarization of unpolarized sources. We have mapped the leakage at GBT across
the beam profile and throughout the passband and find leakage can produce
false linear polarization as large as 10\% and false circular polarization as
large as 30\%. Leakage-produced apparent polarization lacks the
$\lambda^2$ wavelength dependence that we see in the linear polarization data
and cannot produce the 44\% polarization we detect so
we conclude the linear polarization is of astronomical origin
rather than due to leakage.

The detected rotation measure and dispersion measure imply an
electron-weighted average line-of-sight component of the magnetic field of
0.38\,$\mu {\rm G}$, compared to typical large-scale fields of
$\sim10\,\mu {\rm G}$ in
spiral galaxies\cite{2002RvMP...74..775W}.
This field strength is a lower limit for the magnetized region due to
cancellations along the line of sight.
Also, the magnetized region may only weakly overlap the dispersing region and
so electron weighting may not be representative.

The magnetization we detect is likely local to the FRB source rather than in
the Milky Way or the intergalactic medium (IGM). 
Models of Faraday rotation within the
Milky Way predict a contribution of ${\bf RM} =
18(13)\,{\rm rad\,m}^{-2}$
for this line of sight, while the
IGM can contribute at most $ {\bf |RM|} = 6\,{\rm rad\,m}^{-2}$
on a typical line of sight from this
redshift\cite{2015A&A...575A.118O}.

\begin{figure}[tbp]
\centering
\includegraphics[width=9cm]{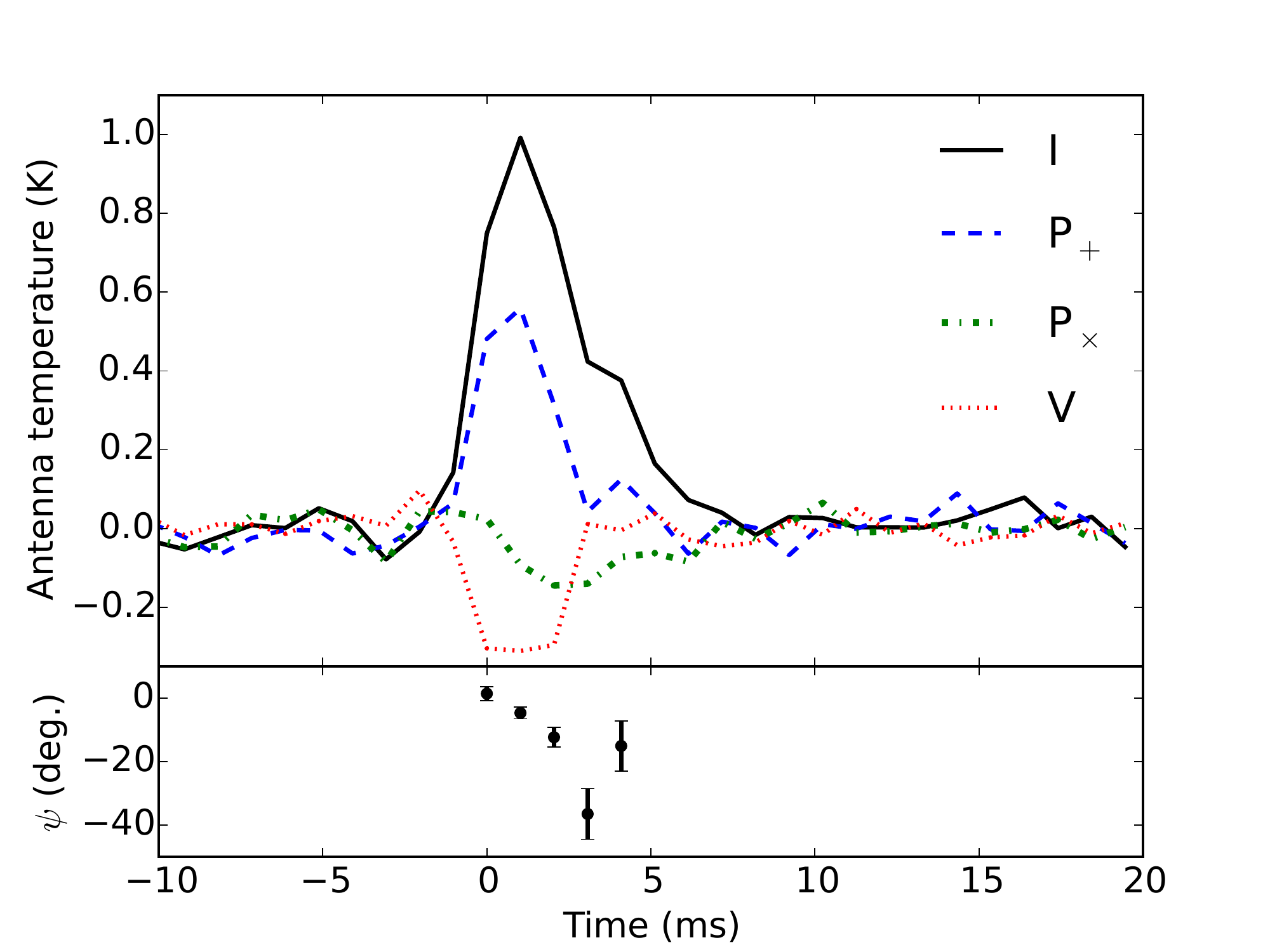} 
    \caption{
\label{fig:pol_profile}
{\bf Polarized pulse profiles averaged over spectral frequency.} Plotted is
total intensity (I), linear polarization (P$_+$ and P$_\times$), and circular
polarization (V, may be instrumental). Before taking the noise-weighted mean
over frequency, the data are scaled to 800\,MHz using the best-fit spectral
index and the linear polarization is rotated to compensate for the best-fit
Faraday rotation.  The linear polarization basis coordinates are aligned with
($+$), and rotated with respect to ($\times$), the mean polarization over time.
Bottom panel shows the polarization angle (where measureable) in these
coordinates. The error bars show the standard error based on simulations.}
\end{figure}

We detect a rotation of the polarization angle over the pulse duration
of $-0.25(5)$\,radian~ms$^{-1}$, shown in 
Figure~\ref{fig:pol_profile}. 
Such rotation of polarization is often
seen in pulsars and is attributed to the changes in the projection of the
magnetic field compared to the line of sight as the neutron star 
rotates\cite{1969ApL.....3..225R}. 

We also detect circular polarization at roughly the 23\% level, but that level of
polarization might be due to instrumental leakage. 
Faraday rotation is
undetectable for circular polarization, so the $\lambda^2$ modulation we use to
identify astronomical linear polarization is not available as a tool to rule out
leakage. 
For these reasons we do not have confidence that the detected circular polarization
is of astronomical origin.

Radio emissions are often scattered: lensing by plasma inhomogeneities creates
multiple propagation paths, with individual delays.  We observe two distinct
scattering time scales in the FRB~110523 data, indicating the presence of two
scattering screens. In five previous FRB detections an exponential tail in the pulse
profile was detected, interpreted as the superposition of delayed versions of the
narrower intrinsic profile. The average scattering time constant for FRB~110523
is $1.66(14)$\,ms at 800\,MHz, with the expected decrease with spectral frequency
as shown in Extended Data Figure~\ref{fig:burst_profiles}.
We also detect scintillation, the variation
of intensity with frequency due to
multi-path interference. We measure a
scintillation de-correlation bandwidth $f_{\rm dc} = 1.2(4)\,{\rm MHz}$ (see
Extended Data Figure~\ref{fig:decorr}), indicating a second source of
scattering with delays of order $1/f_{\rm dc}\sim 1\,\mu{\rm s}$.  This
scintillation is consistent with Galactic expectations for this line of sight.

Scintillation
should only occur if the first scattering screen is unresolved by the second,
and we use this fact to constrain the bulk of the scattering material in the
first screen to lie within 44\,kpc of the source---roughly the scale of a
galaxy (see Methods).
It was previously unknown whether the $\sim$ms scattering observed in FRBs
was due to weakly scattering material broadly distributed along the line
of sight or strong scattering near
the source\cite{2013ApJ...776..125M}, but our detection of scintillation 
eliminates the distributed scattering models.  The observed scattering is 
too strong to be caused by the disks of host
galaxies\cite{2015arXiv150506220K} and therefore
the FRB source must be associated with either a strongly-scattering compact
nebula or with the dense inner regions of its host galaxy.
Either could produce the observed rotation measure, whereas most 
lines of sight through the ISM
of a randomly oriented galactic disk contribute an order of 
magnitude less rotation measure\cite{2015A&A...575A.118O}.

Magnetization and scattering located near the FRB source disfavor models that involve 
collisions of compact objects
such as white dwarfs or neutron stars\cite{2014ApJ...797...70K} 
since these older stellar populations are generally not
associated with compact nebulae nor are they preferentially found near galactic
centres. 
In contrast, a variety of models involving young stellar
populations---including magnetar starquakes, delayed formation of black holes
after core-collapse supernovae, and pulsar giant 
pulses\cite{2014MNRAS.442L...9L,2014A&A...562A.137F,2015arXiv150505535C}---provide
natural explanations for the properties we observe.
Here scattering and magnetization occur in the surrounding young 
supernova remnants or star-forming regions,
and the polarization properties we report are plausible given that these proposed emission
mechanisms involve spinning magnetized compact objects. 
Precise model testing, beyond these general comments, will have to wait 
for more data which will determine if the magnetization and scattering features we
report are generic.

\printbibliography[segment=\therefsegment,check=onlynew,heading=subbibliography]

\begin{addendum}
    \item
        %\red{Acknowledge hippo}
        %
        K.~M.~is supported by the CIFAR Global Scholars Program.
        T.-C.~C.~acknowledges support from MoST grant 103-2112-M-001-002-MY3.
        X.~C.~and Y.-C.~L.~are supported by MOST 863 program 2012AA121701, CAS XDB09000000 and NSFC 11373030.
        P.~T.~T.~acknowledges support from NSF Award 1211781.
        J.~B.~P.~acknowledges support from NSF Award 1211777.
        Computations were performed on the GPC supercomputer at the SciNet HPC
        Consortium.
    \item[Author Contributions]
        K.~M.~integrated the FRB search routines into a software program;
        calibrated and filtered the raw FRB event data; performed scintillation
        analysis; led survey planning; produced Figure~\ref{fig:spectra},
        Figure~\ref{fig:pol_profile}, and
        Extended Data Figure~\ref{fig:decorr}; and
        contributed to model fits to the FRB event, result interpretation,
        beam characterization, observations, data handling,
        and data validation.
        H.-H.~L.~performed the visual search of the search of over 6000 images,
        and discovering the FRB event. He also coproduced Figures~\ref{fig:waterfall} 
        and Extended Data Figure~\ref{fig:burst_profiles}, and contributed to the FRB search
        software program, observations, data handling, and data validation.
        J.~S.~wrote dedispersion and FRB search software routines; performed
        model fits to the FRB event including extracting the dispersion measure,
        rotation measure, scattering tail, and polarization angle swing;
        and contributed to result interpretation.
        C.~J.~A.~contributed to observations, data handling, and data validation.
        T.-C.~C.~contributed to survey planning, observations, and data
        validation.
        X.~C.~contributed to data validation.
        A.~G.~contributed to FRB search algorithm validation.
        M.~J.~contributed to observations and data validation.
        C.-Y.~K.~contributed to observations and data validation.
        Y.-C.~L.~performed scintillation analysis on foreground pulsar and
        contributed to data validation.
        Y.-W.~L.~contributed to polarization leakage
        characterization, calibration methods, and data validation.
        M.~M.~contributed to result interpretation, analysis of the follow-up
        data, scintillation analysis on the foreground pulsar,
        and edited the manuscript.
        U.-L.~P.~carried out Faraday rotation measure synthesis resulting in
        detection of linear polarization and contributed to result
        interpretation, scintillation analysis, survey planning, and data
        validation.
        J.~B.~P.~led manuscript preparation and contributed to result
        interpretation, survey planning, data validation.
        A.~R.~performed survey of archival multi-wavelength catalogues for
        coincident sources, coproduced Figures~\ref{fig:waterfall},
        produced Extended Data Figure~\ref{fig:dmt},
        and added event
        simulation functionality to the FRB search software program.
        P.~T.~T.~contributed to observations and data validation and editing the
        manuscript.
        T.~V.~led observational campaign
        %, produced Extended Data 
        %Figure~\ref{fig:ATNF_dm},
        and contributed to calibration
        methods, survey planning, data handling, and data validation.
        J.~K.~Y.~contributed to data validation.
    \item[Competing Interests] The authors declare that they have no competing
        financial interests.
    \item[Correspondence] Correspondence and requests for
        materials should be addressed to K.~M. (email: kiyo@physics.ubc.ca).
\end{addendum}

% XXX
%\clearpage
%\input{figures}
%\clearpage

\begin{methods}

\subsection{Data and pre-processing}

Our survey was conducted with the GBT linearly-polarized prime-focus 800\,MHz
receiver. For the digital back-end we used the GBT Ultimate Pulsar Processing
Instrument.
The data were collected with the telescope scanning 4 degrees/minute at constant
altitude angle.

To act as a stable flux reference, a broadband noise source injects
power at the feed point, producing a square wave of intensity with period locked at 64 times
the 1.024 ms cadence. In the on-state the noise source increases the total power by approximately 10\%.
The switching noise source must be removed from the data before the search for
transients can proceed.  This is done by accumulating, over the
one-minute scan,
the periodic component with a period of 64 ms. The data
are then normalized to the noise source amplitude in each spectral channel, 
providing an approximate bandpass calibration, and the noise source waveform is
subtracted.  For the search phase this level of calibration is sufficient and no
absolute calibration is applied.

Analysis of the discovered event requires a more rigorous calibration than the
search.  We separately reference the
vertical and horizontal polarization signals to the calibration noise source, 
with the noise source in turn referenced
to a bright unpolarized point source (3C48) scanned 6.5 hours before the event,
providing an on-sky calibration at each frequency and polarization. This
results in an overall absolute flux calibration uncertainly of
9\%\cite{2013ApJ...763L..20M}.
To calibrate the phase of the cross-correlation between the two antenna
polarizations, which we need to measure polarization parameters Stokes U and V, we assume
that the noise source injects the same signal into each with the same phase.
Lab tests of the 800\,MHz receiver verify this assumption except in the two spectral
resonances of the receiver and
in the edges of the band, which we discard. This procedure produces a one
percent polarization calibration at the centre of the beam. The polarization
characteristics off beam centre are described below.

The data contain several spectral channels that are irrecoverably contaminated
by man-made radio frequency interference (RFI), largely due to cell phones.
These are identified by anomalously high variance or skewness relative to other
channels and all data from these channels are discarded.  A total of 3836 out of 4096 channels (93.6\%) pass
the RFI cuts.

Prior to searching the data for FRBs, Galactic and extragalactic synchrotron continuum
emission is removed. Such emission is broadband and varies on much longer
time scales than FRB events and can thus be removed by a variety of algorithms.
For the search, where computational efficiency is a concern, a continuum
template is formed for each 38 second block of data by performing a mean over frequency. 
This template is then
correlated against each spectral channel and the contribution subtracted.

When analysing the discovered event, computational complexity is less of a
concern so we high-pass filter the data on 200\,ms time scales.  The filtering
substantially reduces the variance of the data, and we perform another
iteration of identification of RFI contaminated spectral channels.

\subsection{Searching the data}

To search for FRBs we concentrate the energy
of possible events into a few pixels of an image, using a dedispersion algorithm we developed.
In the array of spectra shown in Figure \ref{fig:waterfall}, the event is
spread out in both time and frequency. We need to remove
this dispersion, aligning the arrival times across frequency, then average over
frequency to produce a time series that has the pulse energy localized.
Since we do not know the dispersion measure a priori we dedisperse over a range of trial values of
DM from 0 to 2000 pc~cm$^{-3}$. The dedispersion process produces a set of frequency-averaged intensities
versus time and DM.  We use a modified tree dedispersion
algorithm\cite{1974A&AS...15..367T}. We developed a recursive program for this algorithm that,
running on a single desktop computer, carries out the dedispersion in 10\% of real time.

After transforming to DM-time space, we need to search each DM for
bursts of unknown duration and unknown profile, which we accomplish 
using a set of boxcar integrals over time, of lengths ranging from 1\,ms to the 
block length of 38 seconds.
Blocks overlap by 8 seconds so events straddling blocks are not missed.
%Could cut this overlap sentence...a fine detail not needed to understand our result
The search algorithm also accumulates noise statistics at each DM for
each boxcar length.  
The procedure is easily parallelized by distributing data files among 
nodes of a large computer array. 
A software package used to search our
data for transient events is publicly
available\cite{burst_search}.

%We find that this algorithm provides 93\% of the
%theoretical maximum signal-to-noise ratio for Gaussian burst profiles. All lengths of
%boxcars are searched in a single pass through the data, so the peak
%detection algorithm runs several times faster than the DM transform
%and orders of magnitude faster than a Gaussian matched-filter
%search.  We note that the 7\% loss in signal to noise ratio applies only to the
%trigger. Once detected, the candidate burst event is treated with more complete models.

The above procedure produced 6496 DM-time plots, which we visually inspected.
We find only one clear FRB candidate---FRB~110523---but the search also 
turned up the previously known pulsars J2139+00 and J0051+0423, roughly in line
with expectations given survey parameters.

We have yet to perform a detailed analysis of the completeness of our search
but taken at face value our single detection implies an all-sky rate of $\sim
5\times10^3$ per day above a fluence of threshold of  $\sim1$\,Jy\,ms,
assuming an effective sky coverage of 0.3\,sq.\,deg.,
in-line with previous estimates.

To provide a set of training templates for the visual search, 
simulated rectangular pulses were added onto a sample of data which included no significant events.
An example of a simulated event is shown in Extended Data Figure \ref{fig:dmt}.
The simulated event shows a characteristic `hourglass' feature 
in the DM-time plots.

\subsection{Modeling the pulse profile and polarization}

We use Markov-Chain Monte Carlo methods to fit a model to the FRB event and
measure its properties.
Throughout the analysis we assume the noise is Gaussian and treat it as
uncorrelated between channels, with per-channel weights estimated from their
variances. This simplification allows us to forgo the time-consuming process of
Fisher matrix statistical analysis.  The assumption is slightly incorrect: the
data are $\chi^2$ distributed with 50 degrees of freedom. We also
find that adjacent frequency channels are actually 2.5\% correlated by the
Fourier transform filter used for spectral channelization. No significant
correlation is detected between more widely separated channels.  We account for
adjacent channel correlation by increasing all errors by 2.5\%.  

To create a model intensity profile for comparison to the data we begin with a Gaussian pulse
profile in time, with width $\sigma$ which is independent of
frequency. This is convolved with a one-sided exponential scattering kernel with a
frequency-dependent duration to yield the normalized pulse profile:
\begin{equation}
    f(\nu, t) = \left[ \frac{1}{\sqrt{2\pi\sigma^2}} \exp(-\frac{t^2}{2\sigma^2}) \right]
        \otimes
        \left[\theta(t) \frac{1}{\tau_\nu} \exp\left(-\frac{t}{\tau_\nu}\right) \right],
\end{equation}
where $\theta(t)$ is the Heaviside step function,
$\tau_\nu = \tau (\nu/\nu_{\rm ref})^{-4}$, the frequency dependence expected for scattering, and
$\tau$ is the scattering time at $\nu_{\rm ref}$.
In the final spectrum we allow for spectral index $\alpha$ of the overall
intensity and delay the pulse for dispersion:
\begin{equation}
    I(\nu,t)=A\left(\frac{\nu}{\nu_{\rm ref}}\right )^{\alpha}
        f(\nu,t-t_{\nu})
\end{equation}
where
$A$ is the burst amplitude at reference frequency $\nu_{\rm ref}$,
$t_{\nu}=t_0+{\rm DM} \times {\rm DM}_0 \left (\nu^{-2}-\nu_{\rm ref}^{-2}\right )$,
${\rm DM}_0=4148.808\,{\rm MHz}^2\,{\rm pc}^{-1}\,{\rm cm}^3\,{\rm s}$,
$\rm DM$ is the dispersion measure of the burst,
and $t_0$ is the burst arrival time at
$\nu_{\rm ref}$.  While in principle the choice of the reference frequency
$\nu_{\rm ref}$ is arbitrary, in practice we find a value of 764.2\,MHz,
near the centre 
of the signal-to-noise weighted band, substantially decorrelates the fit
parameters. This constitutes our base unpolarized model.
Circular polarization is modelled in the same way as total intensity.

Our base linearly polarized model is the same as the unpolarized model with an
added Faraday rotation factor:
\begin{equation}
    [Q+iU](\nu,t)=
    I \exp[2i{\rm RM}(\lambda^2-\lambda_{ref}^2) + i\phi_0],
\end{equation}
where $\rm RM$ is the rotation measure, $\phi_0$ is the polarization angle
at the reference frequency and pulse centre, and $I$ is the model for
intensity given above. 
We find the likelihood surfaces are quite close to Gaussian,
and so the Markov chains converge quickly.  We run an initial short chain to
estimate the covariance matrix, then run 4 chains of length 500,000 steps to
estimate parameters. This approach gives good convergence (the
Gelman-Rubin convergence $r-1$ is typically 0.005).

To search for time dependence of the polarization angle, we extend the
model to allow the polarization angle to vary linearly with time.
We did this fit two ways: 1) apply the phase
gradient to the pre-scattering Gaussian burst and then convolve with the scattering
kernel, and 2) apply the gradient to the scattering-convolved burst
profile.  While the first is more physically natural if the rotation
happens at the burst source before scattering, we find that the second (post-scattering gradient) provides
a significantly better fit (5.4$\sigma$ significance \textit{c.f.} 2.1$\sigma$)
and quote results for this fit. We attribute this to substructure in the
polarized pulse, that is poorly modelled by a Gaussian with linearly changing
polarization angle. We do not have enough signal to noise to further
investigate the substructure but the conclusion that the polarization angle
rotates is robust.

The plasma delay as a function of
frequency is expected to follow a $\nu^{-2}$ power law, scattering time should
have frequency dependence near $\nu^{-4}$, and the Faraday rotation angle should be
proportional to $\nu^{-2}$.  We extend the model used in the Markov
chains to test these predictions, fitting for the dispersion measure index,
scattering index, and rotation measure index. 
All fit parameters are listed in Extended Data Table~\ref{table:FRB_param_main}
with results grouped by independent fits.

To check our analysis software and calibration (especially the polarization sign) 
we performed observations of pulsar B2319+60. The pulsar data were
processed using the FRB pipeline and the Faraday rotation measured from a single
pulse. The rotation measure was determined to be -239.9(4)\,rad\,m$^{-2}$, in
good agreement with the published value\cite{1987MNRAS.224.1073H},
and under this sign convention the FRB's RM
is negative.

\subsection{GBT beam}

During the two second period over which the FRB pulse traverses the bandpass, the
pointing centre of the GBT beam scans 8 arcminutes, which is about half the
FWHM beam width. The pulse intensity increases steeply during the arrival
period likely indicating that the source coordinates moved from the edge of the
GBT beam at the start of the arrival period to a position closer to the beam
centre as lower frequencies arrived. The GBT beam is also wider at lower
frequencies which also contributes to the steep spectral index.  Simulations
indicate that this picture is consistent although, due to the unknown intrinsic
spectral index of the source and unknown impact parameter of the scan relative
to the source, we are unable to use this information to obtain a precise
localization.

It is highly unlikely that the burst entered the telescope through a sidelobe.
Because of its off-axis design, GBT has low sensitivity in its sidelobes.
Simulated models of the 800\,MHz receiver beam show the first sidelobe to
be a ring around the primary beam with radius 0.6\,deg., width 
0.1\,deg.,~and 30\,dB less sensitivity than boresight
(Sivasankaran Srikanth, personal communication, November 2012). The second and
third sidelobes have similar geometry, occurring 0.8 and 1.0\,deg.~from
boresight and suppressed by 37 and 40 dB respectively.
These near sidelobes do
not cover significantly more sky area than the main beam and with their
dramatically lower sensitivity it is unlikely that the lobe would contribute to
the burst detection rate.

Subsequent sidelobes have even less sensitivity but cover more area. They are
ruled out by the observed spectrum of FRB\,110523. The radial width of the 
sidelobes is 0.1\,deg.~and their radial locations are inversely proportional to
observing frequency. As such, if the burst had entered a far sidelobe we would
have observed far more spectral structure; several peaks and nulls. Even the
previously discussed first sidelobe is in tension with the observed spectrum
when accounting for the added spectral structure expected from
$\sim0.1$\,deg.~of scanning during the pulse arrival period. For the first
three sidelobes it is possible that, though an improbable coincidence, the
telescope's scanning could cancel the location spectral dependence of the
sidelobes.
However as previously argued, a source location in the sidelobes is unlikley due to their
combination of low sensitivity and low area.

To determine the polarization properties of the primary beam, we have
performed on-axis and off-axis measurements of the beam using both bright
point sources and pulsars. Such measurements are
crucial for our survey's primary science goal of mapping cosmic structure
through the 21\,cm line. We find that while GBT's off-axis design
reduces sidelobe amplitude it leads to substantial polarization leakage in the
primary beam.  On boresight, the leakage from total intensity to polarization is less than one
percent. Off boresight, leakage peaks at approximatley 0.2\,deg.~in the azimuth direction. Leakage from Stokes I to Q/U is several percent of the
\emph{forward} gain and from Stokes I to V it is as high as 10\%. When
comparing to the gain at that location in the beam instead of the forward gain,
these numbers translate to 10\% leakage to linear polarization and 30\% leakage
to circular. The leakage is only weakly dependent on frequency.
These measurements are in agreement with simulated beam models.

The observed polarization angle rotation over the duration of the pulse
cannot be due to leakage.
The rotation occurs in each frequency bin over 
a few milliseconds,
during which time the GBT beam centre moves just 7 milliarcseconds. 
Gradients of the leakage pattern at such small difference of angle are
much too small to explain the change of polarization angle.
To achieve the signal to noise sufficient to detect the angle swing it is necessary to integrate over 
frequency, introducing the two-second timescale associated with
dispersion delay, but the integrand is composed of millisecond differences of polarization angle, making the
two-second timescale irrelevant.

\subsection{Scintillation}

Since we only see the FRB pulse for a few milliseconds we have no information on 
variation of the flux on longer time scales, and concentrate on quantifying the 
scintillation-induced variation of intensity with frequency by calculating the 
de-correlation bandwidth.
We first form $\delta T(\nu) \equiv T(\nu) / T_{\rm
smooth}(\nu)$ where $T_{\rm smooth}(\nu)$ is the power law fit to the spectrum,
accounting for the intrinsic spectrum of the event as well as the frequency dependence and motion of the telescope beam.  We then form the correlation function
\begin{equation}
    \xi(\Delta\nu) \equiv \langle \delta T(\nu) \delta T(\nu + \Delta\nu)
    \rangle_{\nu}.
\end{equation}
This correlation function is estimated from the observed spectrum and is shown in
Extended Data Figure~\ref{fig:decorr}.

To estimate the de-correlation bandwidth, $f_{\rm dc}$, from the observed correlation function,
we fit to the Fourier
transform of an exponential scattering function\cite[Chapter~4]{2004hpa..book.....L}:
\begin{equation}
    \xi_{\rm model}(\Delta\nu) = \frac{m}{f_{\rm dc}^2 + \Delta\nu^2}.
\end{equation}
This fit yields $f_{\rm dc} = 1.2(4)\,\rm{MHz}$ and $m=0.26(8)$. The errors on the
measurement of the correlation function depend on the underlying statistics of
the scintillation, which are both non-Gaussian and
model-dependent\cite{2012ApJ...755..179J}.
We estimate the errors in Extended Data Figure~\ref{fig:decorr} through
simulations, with errors on fit parameters subsequently expanded
to account for modelling uncertainties.

\subsection{Two-screen model for scintillation and scattering}

The observed scintillation de-correlation bandwidth is comparable
to that observed for Galactic pulsar
J2139+00, less than two degrees away from FRB~110523 on the sky and at a
distance of 3\,kpc based on its dispersion measure\cite{1993ApJ...411..674T},
indicating the
scintillation arises from the Galactic interstellar medium.

A familiar form of scintillation in optical astronomy is the twinkling of stars.
Optical scintillation is due to turbulence in the atmosphere and is commonly 
modelled by projecting the optical medium onto a screen above the telescope 
with micro-images appearing in the plane of this screen.
For stars, rapid variation of flux with frequency are seen 
because stars have angular size small enough that light emitted from opposite
edges of the stellar disk has path length difference less than a wavelength. 
Stars are said to be unresolved by the scintillation screen, meaning that they are
indistinguishable from point sources. The multipath
interference changes with time because of turbulent motions in the atmosphere.
%In optical astronomy the micro-images are called speckles.
Planets, in contrast to stars, have angular size resolved by the screen,
so the flux variations are typically a small fraction of the total flux. 
For similar reasons, among radio sources,
pulsars often show scintillation, while the much larger extragalactic radio sources do not.
At radio wavelengths scattering occurs in the intervening plasma rather than the atmosphere.
%Scintillation of pulsars in the Milky Way is due to plasma density fluctuations\cite{2014MNRAS.442.3338P} with lens image separations of afew milliarcsecconds, which translates to several AU at a characteristic distance of 1 kpc.

To model scintillation and scattering for FRB 110523 we project the intervening 
material onto two screens, representing the material in the Milky Way and in the
host galaxy, respectively. We use two screens because the scintillation and 
scattering have very different time scales, which precludes modelling with a single screen.
As with optical scintillation each screen produces a halo of micro-images,
which can be considered scattering sites.
Propagation via a micro-image at the edge of a halo requires a longer propagation 
time from source to observer than micro-images near the centre. In our model the
delays associated with the Galactic screen produce the micro-second scintillation
path differences, while the host screen path differences produce the 1.6 ms
exponential tail of the pulse profile.

In our two-screen model the presence of strong scintillation indicates that
the host screen is unresolved by the Galactic screen, and this allows an 
estimate of the host screen position. We assume the position of the Galactic
screen is the characteristic thickness of the ionized Galactic plane
$D$=1~kpc. The angular size  the Galactic screen is then given
by $\theta = \sqrt{2 c\tau /D} \sim 1$~mas and the resolving power of the Galactic
screen is $\rho=\lambda/(\theta D)\sim$ 600~nas.
The scintillation would be washed out if the host screen exceeded this angular size.
This small angular size combined with the 1.6 ms scattering time places 
the host scattering screen within 44\,kpc of the source, assuming the maximum
source distance of $\sim1$\,Gpc (constrained by the observed
dispersion measure).

To further test our scintillation and scattering model, we compared the
scintillation of the
main pulse to the scintillation in the scattering tail by cross-correlating the intensity spectrum
early in the pulse to the spectrum late in the pulse. To
obtain the early pulse spectrum, we use a filter matched to the Gaussian part
of the profile with no scattering tail. For the late part we use a filter
matched to the tail beginning 3\,ms into the pulse. The cross de-correlation
bandwidth is $f_{\rm dc} = 1.3(5)$ compared to $f_{\rm dc} = 1.1(6)\,\rm{MHz}$ and
$f_{\rm dc} = 1.0(4)\,\rm{MHz}$ for the early and late pulse respectively.
Correlation amplitudes are $m=0.30(9)$, $m=0.18(8)$, and
$m=0.47(13)$ for the cross correlation, early, and late pulse respectively.
These are all consistent with the level of scintillation measured for the full
pulse, indicating that the most direct path and scattering-delayed micro-images
share a common scintillation-induced spectrum. The scintillation source is
therefore separate from the source of the scattering tail, and we place them in
the Milky Way and host galaxy respectively.

\subsection{Follow-up observations}

We carried out observations at the position of FRB 110523 from 700~MHz to 900
MHz at three separate epochs on MJDs 57134, 57135, and 57157 for durations
of 1.8 hrs, 1.8 hrs, and 3 hrs, respectively. We detected no bursts with
DMs in the range of 0 to 5000\,pc\,cm$^{-3}$ with
significance greater than 6 sigma. We also performed a periodicity search on
the data, and detected no pulsar candidates. The estimated limiting flux
density of this search, assuming a pulsar duty cycle of 10\%, was 0.04~mJy.

\subsection{Counterpart sources}

To identify possible optical counterpart source candidates we searched the
Sloan Digital Sky Survey (DR12) catalogues\cite{2015ApJS..219...12A}
throughout a region centred on the
position of the radio beam at the time the pulse arrived at 700 MHz.
The beam size of the GBT is 15 arcminutes FWHM, 
but we expanded the search area to 30 arcminutes diameter to account 
for a source lying outside the FWHM beam area.
Within this field there are 70 objects identified as galaxies in the
catalogue of which 40 are listed as having redshift less than 0.5.
The 100\% galactic completeness limit of SDSS photometry\cite{2001AJ....122.1104Y} is
r-band magnitude 21. As such, all Milky Way-like galaxies are
included for $z < 0.28$, assuming an absolute magnitude $M_{\rm r} \approx -19.86$.

No X-ray or gamma-ray sources are listed in the
NASA/IPAC Extragalactic Database in this region.

\subsection{Data availability}

The raw data used in this publication
%will be made available online upon acceptance of this manuscript.
are available at \url{http://www.cita.utoronto.ca/~kiyo/release/FRB110523}.

\subsection{Code availability}

The code used to search the data archive for FRB events is available at
\url{https://github.com/kiyo-masui/burst_search}.

The code used to analyse the discovered FRB is available at
\url{https://github.com/kiyo-masui/FRB110523_analysis}.

\end{methods}

\printbibliography[segment=\therefsegment,check=onlynew,heading=subbibliography]

\clearpage
\begin{extended-data}

\begin{table}[hbp]
\begin{center}
{
%\tabulinesep=10mm
\extrarowsep=-0.1ex
\sffamily\sansmath%
% XXX
%\footnotesize
\newlength{\sectsep}
\newlength{\valsep}
\setlength{\valsep}{0.3ex}
\setlength{\sectsep}{1.2ex}
\begin{tabu}{lc}
\\
\hline\\[\valsep]
Barycentric $\nu=\infty$ arrival (MJD) & $55704.62939511$\\[\valsep]
GBT boresight at 900\,MHz arrival & $\rm RA=21^h45^m31^s$\\
                                    & $\rm Dec=-00^d15^m23^s$\\
                                    & $l=56.0795^\circ$\\
                                    & $b=-37.9435^\circ$\\[\valsep]
GBT boresight at 700\,MHz arrival & $\rm RA=21^h45^m12^s$\\
                                    & $\rm Dec=-00^d09^m37^s$\\
                                    & $l=56.1215^\circ$\\
                                    & $b=-37.8234^\circ$\\[\sectsep]
%\hline
Dispersion measure (\,pc\,cm$^{-3}$) & $623.30 (6)$\\[\valsep]
Fluence at 800\,MHz (K\,ms) & $3.79(15)$\\[\valsep]
Spectral index & $-7.8 (4) $\\[\valsep]
Unscattered pulse FWHM (ms) & $1.73 (17)$\\[\valsep]
Scattering time at 800\,MHz (ms) & $1.66 (14)$\\[\sectsep]
%\hline
Linear polarization fraction (\%) & $44 (3)$\\[\valsep]
Rotation measure (rad m$^{-2}$) & $-186.1 (1.4)$\\[\sectsep]
%\hline
Polarization rotation rate (rad ms$^{-1}$) & $ - 0.25 (5)$\\[\sectsep]
%\hline
Dispersion measure index &$-1.998 (3)$\\[\sectsep]
%\hline
Scattering index & $-3.6 (1.4)$\\[\sectsep]
%\hline
Faraday rotation index &$-1.7 (2)$\\[\sectsep]
\hline
\\
\end{tabu}

\caption{{\bf FRB~110523 parameters.} Arrival time and astrometric parameters
as well as parameters for fits of the base unpolarized, base polarized,
and extended models to antenna temperture data.
The steep spectral index we observe is attributed to beam
effects. Statistical uncertainties enclose the 68\% confidence
interval of the measurement.
}
\label{table:FRB_param_main}
}
\end{center}
\end{table}

\begin{figure}[hbp]
\includegraphics[width=18cm]{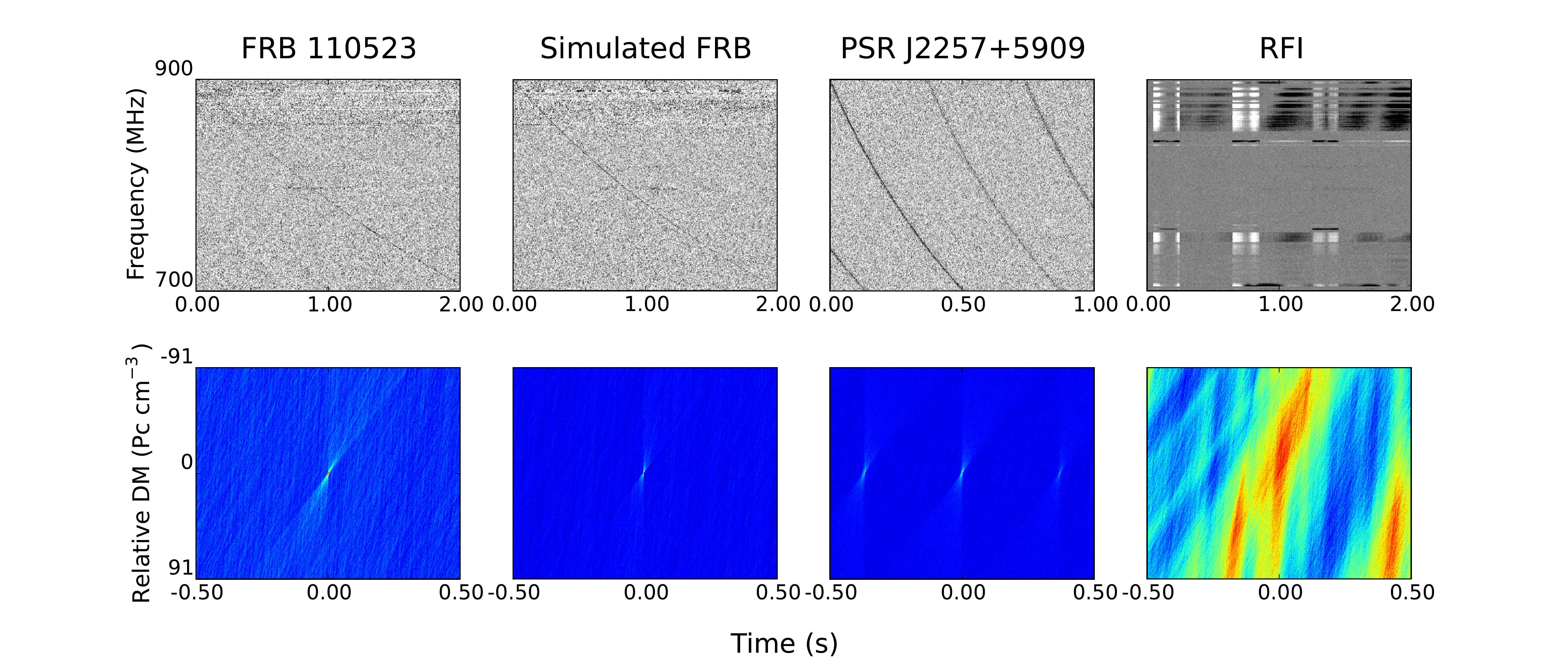} 
    \caption{
\label{fig:dmt}
{\bf Events in Frequency--time and Dispersion Measure--time space.}
From left to right: Data for FRB~110523; a simulated FRB; a known pulsar PSR J2257+5909; 
man-made Radio Frequency Interference.
Brightness temperature is shown in frequency-time space (upper panels) and the same data in 
dispersion measure-time space (lower panels).
Relative dispersion measure is is the difference between the DM and the event DM; event DMs are 622.8, 610.3, 151.0, and 1132.1 pc cm$^{-3}$,
respectively from left to right. The time axes of the frequency-time plots show time relative to the DM-t zero time.
The colour scale in the lower panels represents broadband flux with 
red showing a bright source.
}
\end{figure}

\begin{figure}[hbp]
\centering
\includegraphics[width=9cm]{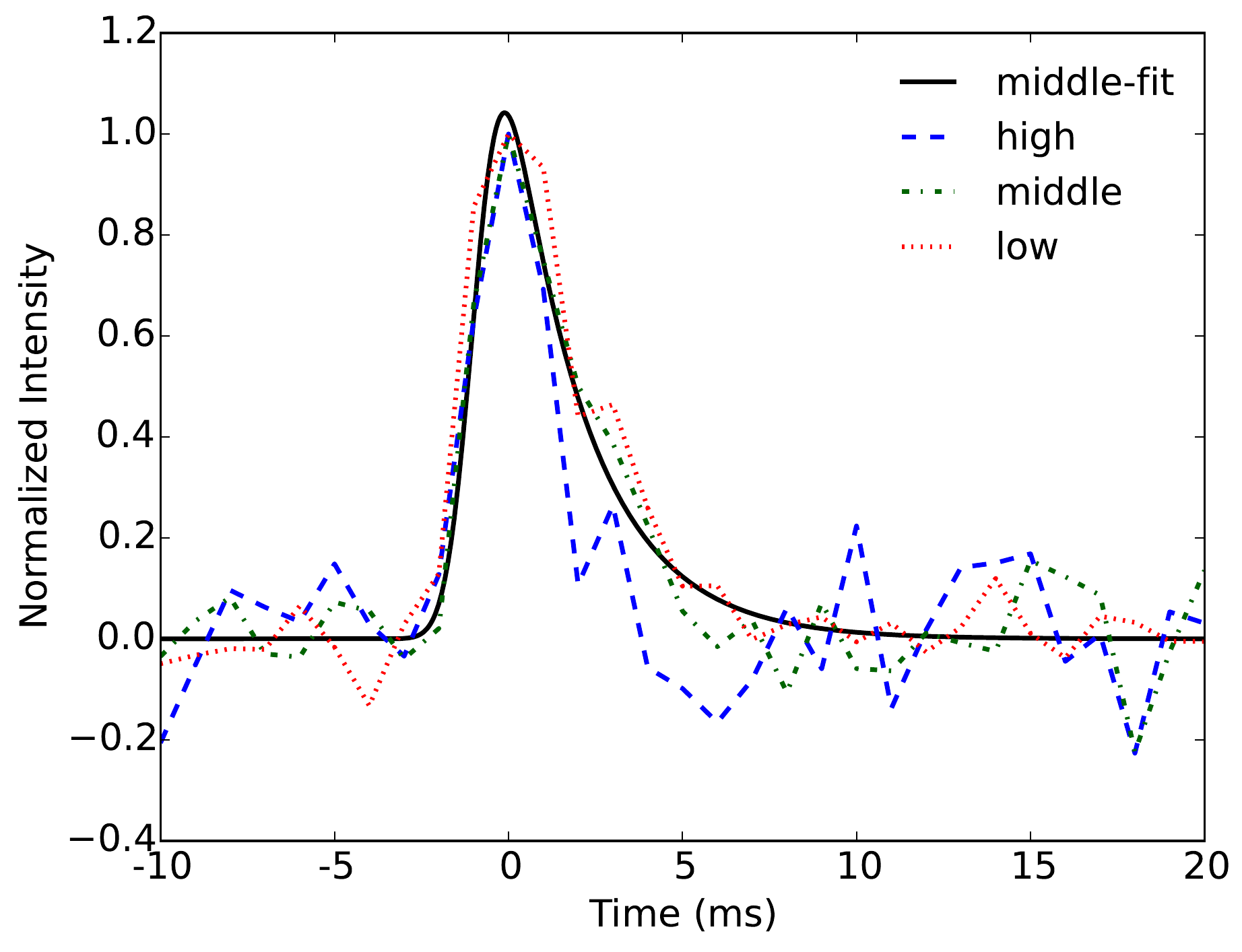}
\caption{
\label{fig:burst_profiles}
{\bf Pulse profiles for FRB~110523 in three sub-bands.} Each sub-band has
width 66 MHz. The pulse width decreases with frequency (at 2.6-sigma
significance), consistent with models
of scattering in the interstellar medium.  Also shown in black is the best-fit
model profile for the middle band.}
\end{figure}

\begin{figure}[hbp]
\centering
\includegraphics[width=9cm]{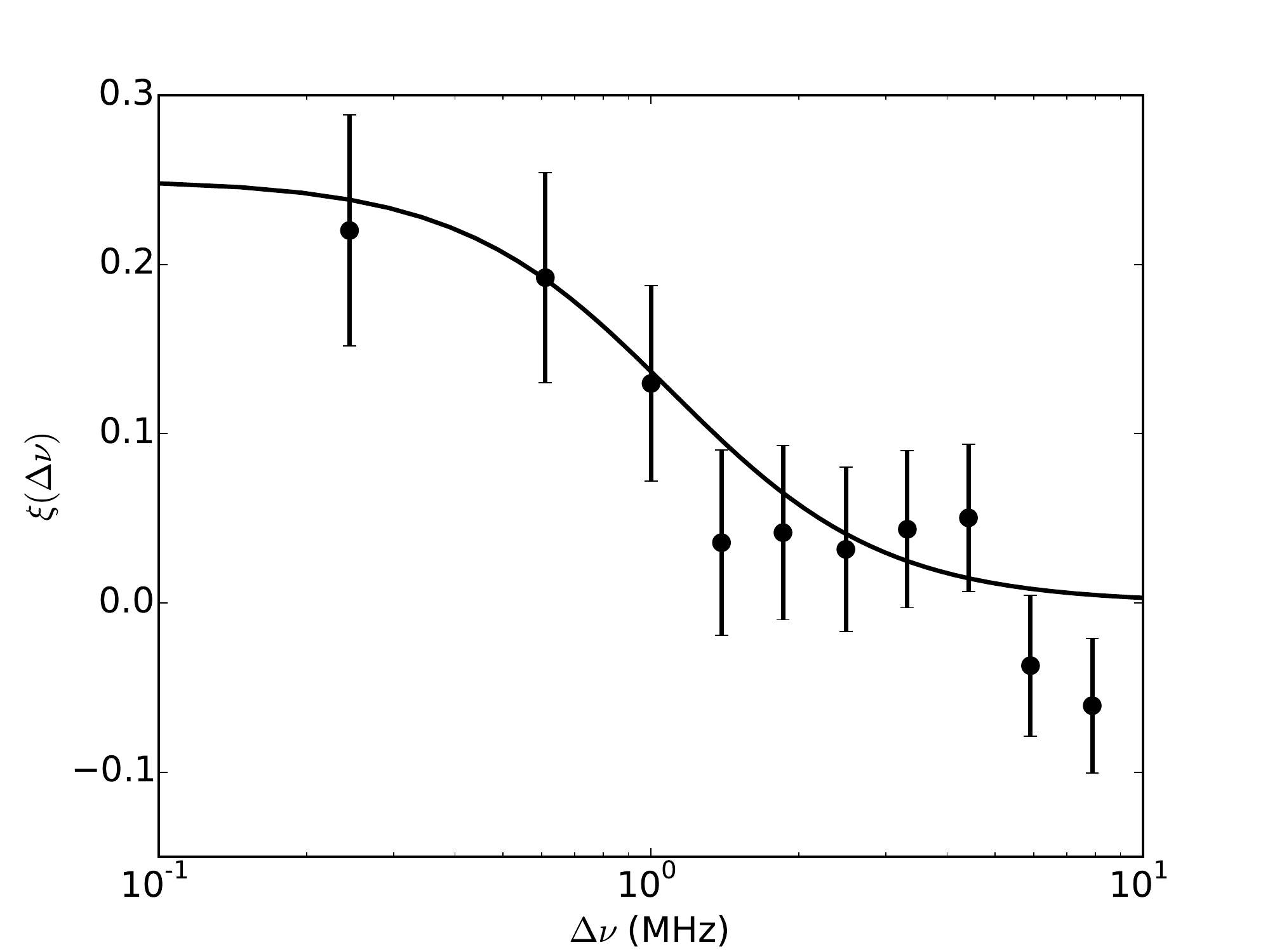} 
    \caption{
\label{fig:decorr}
{\bf Spectral brightness correlation function of FRB~110523.} The intensity spectrum has
structure that is correlated for frequency separations less than $f_{\rm dc} = 1.2\,\rm{MHz}$.
Error bars are standard errors estimated from simulations and are
correlated.}
\end{figure}

\end{extended-data}

\clearpage

\end{document}